\newcommand{\CLASSINPUTtoptextmargin}{0.75in}
\newcommand{\CLASSINPUTbottomtextmargin}{1.05in}
\newcommand{\CLASSINPUTinnersidemargin}{0.64in}
\newcommand{\CLASSINPUToutersidemargin}{0.64in}
\documentclass[conference]{IEEEtran}
\IEEEoverridecommandlockouts
\usepackage{cite}
\usepackage{amsmath,amssymb,amsfonts}
\usepackage{algorithmic}
\usepackage{graphicx}
\usepackage{textcomp}
\usepackage{xcolor}
\usepackage{url}  
\usepackage[draft,bookmarks=false]{hyperref}
\def\BibTeX{{\rm B\kern-.05em{\sc i\kern-.025em b}\kern-.08em
    T\kern-.1667em\lower.7ex\hbox{E}\kern-.125emX}}
\begin{document}

\title{Interface on demand: Towards AI native Control interfaces for 6G\\
{}
}

\author{\IEEEauthorblockN{Abhishek Dandekar}
\IEEEauthorblockA{\textit{TU Berlin} \\
Berlin, Germany \\
a.dandekar@tu-berlin.de}
\and
\IEEEauthorblockN{Prashiddha D. Thapa}
\IEEEauthorblockA{\textit{Fraunhofer HHI} \\
Berlin, Germany \\
prashiddha.dhoj.thapa@hhi.fraunhofer.de}
\and
\IEEEauthorblockN{Ashrafur Rahman}
\IEEEauthorblockA{\textit{TU Berlin} \\
Berlin, Germany \\
a.rahman@tu-berlin.de}
\and
\IEEEauthorblockN{Julius Schulz-Zander}
\hspace{20cm}
\IEEEauthorblockA{\textit{Fraunhofer HHI} \\
Berlin, Germany \\
julius.schulz-zander@hhi.fraunhofer.de}
}
\IEEEpubid{\makebox[\columnwidth]{~\copyright IEEE International Conference on Communications 2025\hfill}%
\hspace{\columnsep}\makebox[\columnwidth]{}}

\maketitle

\begin{abstract}
Traditional standardized network interfaces face significant limitations, including vendor-specific incompatibilities, rigid design assumptions, and lack of adaptability for new functionalities. We propose a multi-agent framework leveraging large language models (LLMs) to generate control interfaces on demand between network functions (NFs). This includes a matching agent, which aligns required control functionalities with NF capabilities, and a code-generation agent, which generates the necessary API server for interface realization. We validate our approach using simulated multi-vendor gNB and WLAN AP environments. The performance evaluations highlight the trade-offs between cost and latency across LLMs for interface generation tasks. Our work sets the foundation for AI-native dynamic control interface generation, paving the way for enhanced interoperability and adaptability in future mobile networks.

\end{abstract}

\begin{IEEEkeywords}
AI native, interfaces, 6G, agents, LLM
\end{IEEEkeywords}

\section{Introduction}
Mobile networks have evolved over time from being monolithic and tightly coupled to being more flexible. This started with the emergence of Software-Defined Networks (SDN) which separated the network control plane from the network data plane. This meant that instead of tightly coupled software and hardware, it was possible to run the software separately and remotely from the hardware. It paved the way for disaggregated networks. In disaggregated networks, instead of combining all functionality in a single Network Function (NF), it is split into multiple NFs. For example, in 5G networks, a single gNB can be split into Radio Unit (RU), Distributed Unit (DU), Control Unit - Control Plane (CU-CP), and Control Unit - User Plane (CU-UP)\cite{23501}. When a monolithic NF is split into multiple NFs, they require new interfaces to communicate between them. These interfaces can be control interfaces (e.g., F1-C) or user data interfaces (e.g., F1-U). The entire mobile network can be seen as a set of NFs that communicate and coordinate with each other over these interfaces.

In order to operate networks with disaggregated and distributed network functions, it is essential to define these network interfaces in advance. This is done by various standardization organizations (SDOs). However, using such predefined standardized interfaces also creates various issues.
\begin{itemize}
    \item Vendors of all NFs may not be able to conform with these standardized interfaces, thus making multi-vendor deployment challenging.
    \item Standardized network interfaces are often specific to a particular Radio Access Technology (RAT) which means the NFs belonging to one RAT cannot communicate with another RAT.
    \item These interfaces are inflexible and designed assuming that the functionality of the NFs remains fixed. If a new functionality is to be added to NF, it cannot use the predefined network interface. 
\end{itemize}

In order to address these challenges, we propose a framework where we can create a control interface between any two NFs on demand. We achieve this using a multi-agent framework based on Large Language Models (LLMs). 

In Section II, we provide a brief background on LLMs, followed by an overview of our framework in Section III. Section IV explores one of the use cases realized using our framework, and Section V describes its implementation and evaluation. In Section VI, we discuss the current limitations of our framework. In section VII, we provide relevant research work and in the final section we conclude our work.

\section{Background}
\subsection{Large Language Models}
Large language models are machine learning models trained on vast quantities of textual data and are used to generate text in natural language\cite{NIPS2017_3f5ee243}. These models are non-deterministic in nature and operate by predicting the next word in a sequence, iteratively constructing coherent and contextually relevant text. In order to generate an output these model needs to be provided with a set of instructions which are termed as \textit{prompt}. A \textit{context window} refers to the amount of text or data that the LLM can process and generate at a single time. LLMs have a finite context window hence the amount of text that could be generated and provided in a prompt is limited. In cases where it is necessary to provide a larger amount of text, techniques like Retrieval Augmented Generation (RAG)\cite{RAGneurips} are used.

Using RAG, the LLM utilizes retrieved information to generate outputs that are both factually grounded and also highly relevant and accurate. This approach significantly reduces the chances of producing incorrect or irrelevant responses while offering additional benefits such as enhanced adaptability across domains, real-time integration of the latest data, and improved user trust through source attribution. RAG ensures minimal hallucination and generates precise, reliable, and tailored outputs.


\subsection{Agentic LLMs}

An agent\cite{Wooldridge_Jennings_1995} can be seen as a set of software pipelines built on top of the LLM engine which allows access to external information, cross-checking, and reasoning. It is called an agent because it has \textit{agency}, which is the ability act on its own without requiring constant input from a human. An agent can be specialized to perform specific tasks. In a multiagent system, multiple such specialized agents act together to achieve a certain goal. For example, an itinerary agent can create an itinerary for a trip based on user requirements. This agent can then forward the itinerary to a ticket booking agent which can fetch information from various airlines and book the most suitable ticket. Such multi-agent frameworks can be used to reach a certain goal from a set of requirements.

\section{Interface Generation framework}

This section provides an overview of the key steps required for generating a control interface between two NFs. 

\begin{figure}[h]
    \centering
    \includegraphics[scale=0.3]{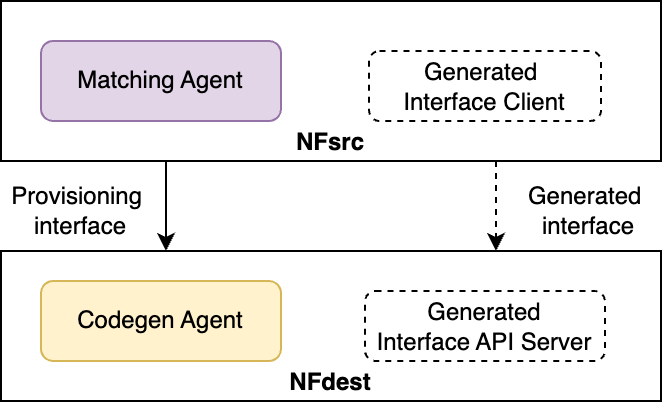} 
    \caption{Multi-agent framework}\label{fig:main}
\end{figure}

Consider two sets of NFs, a source NF (\textit{NFsrc}) and a destination NF (\textit{NFdest}) as shown in Figure~\ref{fig:main}. Let's assume \textit{NFsrc}  needs to generate a control interface towards \textit{NFdest} in order to control its functionality. It uses two LLM based agents, Matching agent in \textit{NFsrc} and Codegen agent in \textit{NFdest} to generate this interface. Before the interface generation process can begin, \textit{NFsrc} needs to discover \textit{NFdest} using appropriate methods (e.g. DNS based discovery, static configuration, etc.). Subsequently, it needs to establish trust with \textit{NFdest} using suitable mechanisms. Then a provisioning interface is setup between \textit{NFsrc} and \textit{NFdest} based on a predefined port. This interface is used by the respective agents to communicate with each other.

\subsection{Matching agent}
The initial step in generating interfaces is to meet two types of compatibility requirements. Firstly, both NFs need to be functionally compatible, which means that the target \textit{NFdest} should have the capability to support the functionality required by the \textit{NFsrc} (e.g. ability to change the transmission channel). Secondly, they need to be semantically compatible, this means that the \textit{NFdest} needs to correctly understand the incoming control message from \textit{NFsrc} irrespective of  varying function names. For example, \textit{NFdest} needs to understand that \textit{set\_channel} and \textit{setchn} both refer to the same functionality of setting the channel number. Both NFs also need to have compatible encoding/decoding schemes for  control messages (e.g. protocol buffer, flatbuffer, etc.).

\begin{figure}[h]
    \centering
    \includegraphics[scale=0.23]{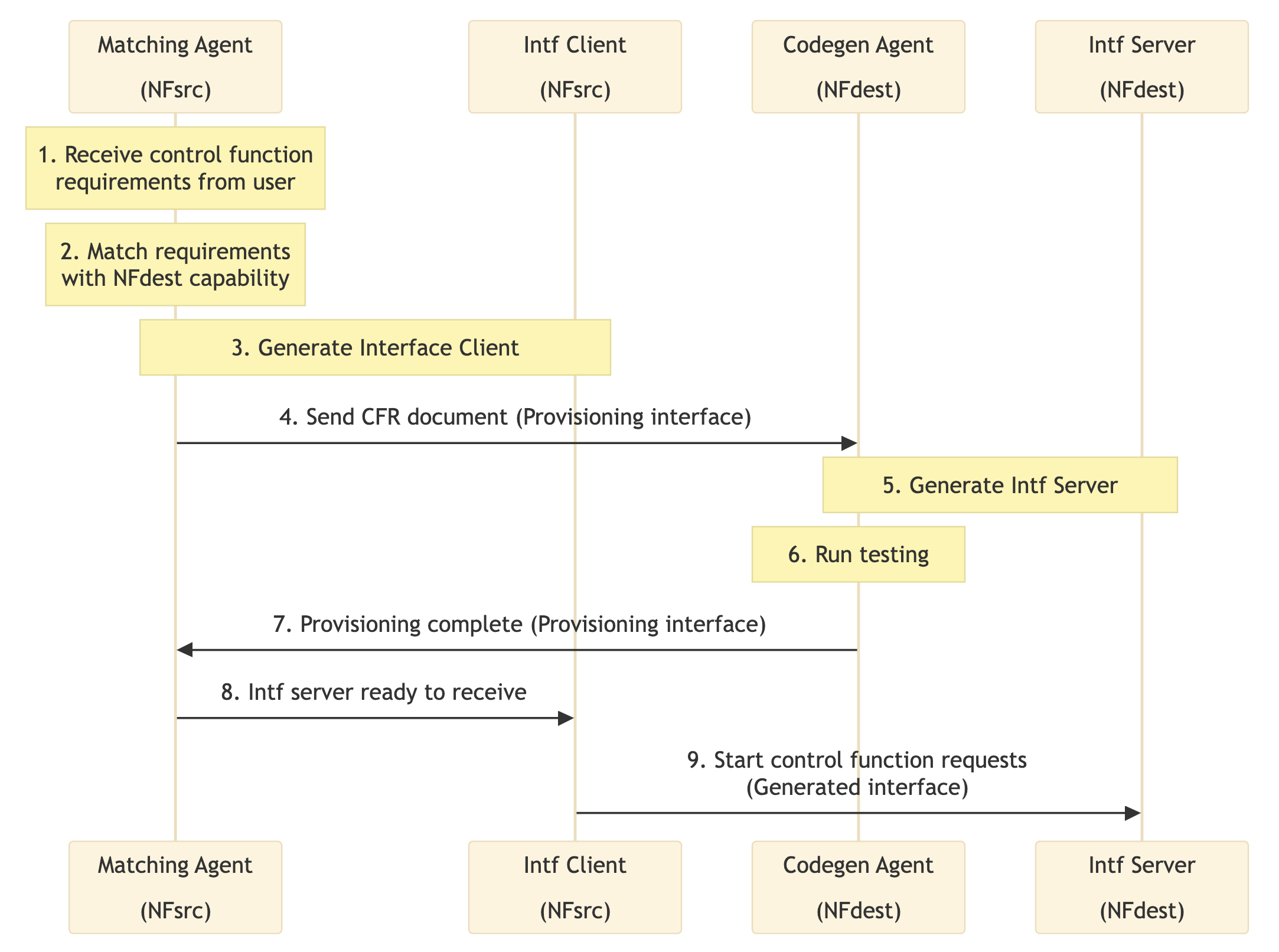} 
    \caption{Interface generation workflow}\label{fig:seq}
\end{figure}

When a user intends to send certain control functions from \textit{NFsrc} to \textit{NFdest}, it needs a mechanism to check if the required functionality is supported by the \textit{NFdest}. To achieve this the user sends control function requirements to the matching agent (Fig.~\ref{fig:seq} step 1).  The matching agent is an LLM based agent which allows the user to find whether the required control functions are supported and to identify the required input parameters. This is achieved by matching the required control function description from the user against the \textit{NFdest} capability document (Fig.~\ref{fig:seq} step 2). The \textit{NFdest} capability document details the capabilities of the \textit{NFdest}. It can either be provided from an online repository of capability documents or directly from the \textit{NFdest} itself.

If an exact match is not found among the NF control functions, the closest match is selected, which can later be \textit{augmented} by the codegen agent. If the agent cannot find any similar function, it informs the user that the NF is unable to support required control functions. Note that this matching is done in terms of available functionality and not in terms of the exact API of the \textit{NFdest}. This allows NF vendors to not expose their API externally.

If \textit{NFdest} supports the required functionality then a control function requirements (CFR) document is created. This document contains a list of required control functions and their matching functions from capability document along with the required encoding scheme.
It can include the required control function name, input parameters, output parameters, description of the control function and matched control function. It also includes details like parameter data types, units, etc. For example:\\
\\
\textit{func setpower (radioID string, pow string dBm)(response boolean): \textless description \textgreater:\textless matched function\textgreater
}\\

The matching agent then generates an interface client which allows to send and receive control function messages to the \textit{NFdest} (Fig.~\ref{fig:seq} step 3). The CFR document is then sent to the \textit{NFdest} 
 (Fig.~\ref{fig:seq} step 4). This document can be sent using a REST POST query over the provisioning interface.


\subsection{Codegen agent}

The CFR document sent from the matching agent is received by the codegen agent on \textit{NFdest}. This agent is tasked with creating an interface API server app that can decode and execute the control functions provided in the CFR document. The codegen agent generates this interface API server using knowledge from an internal API document which is present on the \textit{NFdest} (Fig.~\ref{fig:seq} step 5). The internal API document is a developer document specific to the vendor, detailing how to use the internal API to trigger the control functions supported by \textit{NFdest}. Both the internal API and the API documentation can be proprietary and need not be exposed externally. The generated interface API server performs the following functions:
\begin{itemize}
    \item Decode the incoming control message sent by the interface client from \textit{NFsrc}
    \item If required adapt the unit or data type of control function parameters
    \item Trigger the corresponding internal APIs 
    \item If required adapt the units and data types of parameters returned from the internal API 
    \item Encode and send this information back to \textit{NFsrc} interface client
\end{itemize}

In addition to this, the codegen agent can also generate augmented functions. This happens when the required control function does not have a direct matching function in the internal API. The codegen agent generates a wrapper function around the  internal API function which is most similar to support the additional functionality required by the control function.

Once the code generation is done, the agent generates a test interface client code with dummy parameter values to check the validity of the generated interface server. This test code may also be provisioned by the \textit{NFsrc}. If the interface server code throws errors in this test, this error is fed back to the LLM to regenerate this code. The agent can do it multiple times until the correct version of the code is generated (Fig.~\ref{fig:seq} step 6). Once the correct code is generated the codegen agent sends a provisioning complete message to the \textit{NFsrc} to indicate that the new control interface is ready (Fig.~\ref{fig:seq} step 7).  The matching agent then triggers interface client to initiate a connection (Fig.~\ref{fig:seq} step 8). After this step, the interface client initiates a connection with the interface server and starts sending control function requests (Fig.~\ref{fig:seq} step 9).

\section{Use case}

In this section, we provide an illustrative example of how the multiagent framework described in previous sections can be used to generate interface on demand for controlling IEEE
802.11(WLAN) and 5G NFs on demand. O-RAN is a framework that disaggregated the 5G gNB-DU into O-DU, and O-RU. As per the principles of SDN, it also separates the control plane from 5G gNB and places it in two distinct controllers\cite{polese2023understanding}. The Non-Real-Time RIC (Non-RT RIC) controls tasks with higher latency (\textgreater1s) while the Near-Real-Time RIC (Near-RT RIC) handles control tasks requiring lower latency (1s\textless). The Non-RT RIC sends control messages to the DU and the CU using the O1 control interface while the Near-RT RIC controls these NFs over the E2 control interface. Both these interfaces have been standardized by the O-RAN alliance. However, they exhibit the  following issues:

\begin{itemize}
    \item Managing a multi-vendor O-RAN deployment is challenging.
    Even though the DU/CU might use the standardized E2 interface, it may not be compatible with the near RT-RIC. This is due to divergence in implementation, and especially in the encoding schemes. A good example of this is the OSC-RIC~\cite{osc-RIC-2024-12-22} and  FlexRIC~\cite{mosaic5g}. As they have different E2 implementations, each DU/CU needs to have an OSC-RIC E2 agent and a FlexRIC E2 agent in order to communicate with the near-RT RIC. In a nutshell, the controllers and DU/CU are functionally compatible but they are semantically incompatible.
    \item Making changes to predefined, standardized interfaces like E2 requires changing the service module (SM) and also the low-level code of the DU/CU. This limits innovation as it slows the development cycles.
    \item  3GPP allows for using non-3GPP technologies like WLAN with 5G. However, SDOs like O-RAN do not specify a control interface towards WLAN, because of which joint control of 3GPP 5G and WLAN is not possible. These networks currently do not adapt to each other to achieve optimal resource utilization. In order to have joint control, we need a Multi-RAT RIC which can control both gNB and AP. With multi-RAT RIC, the RIC applications can coordinate and control both 5G and WLAN networks \cite{morph}. For instance, if the WLAN link provides more reliable latency and the 5G link offers higher throughput, resource allocation can be adjusted to prioritize latency on WLAN link and throughput on 5G link. 

\end{itemize}

\begin{figure}[h]
    \centering
    \includegraphics[scale=0.24]{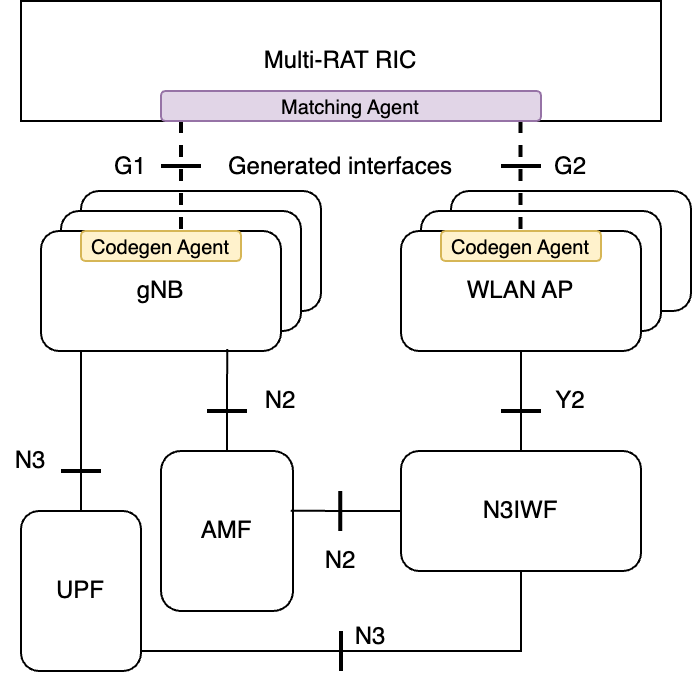} 
    \caption{Joint control for WLAN and 5G}\label{fig:ucase}
\end{figure}

Figure~\ref{fig:ucase} shows a scenario for joint deployment of 5G and WLAN. Note that there can be multiple APs and gNBs from various vendors. WLAN communicates with the 5G core network through N3IWF (N3 Inter working function) as defined by 3GPP\cite{23501}. The core network treats the WLAN Access Point (AP) as transparent and hence exerts no control over it. In order to control WLAN AP, multi-RAT RIC generates G2 and it generates G1 for controlling gNB. 

 Initially, the RIC discovers and sets up a trusted connection with all the target gNBs and APs. The user then provides control functions requirements to the matching agent. The agent then generates the CFR document with the matched gNB or AP control functions. This CFR document is then sent to gNB and AP. This information is received by the codegen agents on each NFs. These agents generate the code required for creating the interface API server. Note that the API server is unique for every vendor. The agent tests the code until the correct version is generated. Once this is successful it then sends a provisioning complete message back to RIC. The RIC then triggers the interface client to start sending control messages.

\section{Implementation and Performance}

In order to test our use case, we use a custom multi-RAT RIC implementation with an embedded matching agent as shown in Figure~\ref{fig:ucase}. We simulate gNB and WLAN AP using containers. As commercial vendors do not divulge their internal API, we created a set of custom internal APIs for both gNB and AP which can trigger simulated control functions. For example, \textit{getRateStats} and \textit{releaseUE}. Our API simulates 30 control actions each for gNB and AP.
In order to simulate a multivendor scenario, we created multiple sets of APIs that are functionally similar but semantically different (e.g. varying function names, data types and units). We test using 5 simulated multivendor gNBs and 5 simulated multivendor APs. 

For both agents, we use a RAG pipeline in combination with two LLMs:  GPT-4o and Llama3.3-70B. The GPT-4o model runs on OpenAI servers, while the Llama3.3 model runs on a server with Nvidia Tesla V100 (128GB VRAM). We use bge-small-en-v1.5 as the embedding model with Llama3.3 and text-embedding-ada-002 as an embedding model for GPT-4o.

The RAG pipeline consists of three stages- document indexing, query-based retrieval, and response generation. In the document indexing stage, the input text document is tokenized to create embeddings. These embeddings are stored in a vector database for later retrieval. During the query-based retrieval stage, the vector database is searched for text embeddings most relevant to the LLM input query. The top 3 most relevant embeddings are fetched and are passed to the LLM. In the response generation phase, the LLM generates a reply to the input query based on the retrieved text embeddings. In case of the matching agent, the NF capability document and the control function requirements from user are used as input documents for the RAG pipeline. 
For the codegen agent, the CFR document along with vendor API documentation is used as input for its RAG pipeline. 
We use custom prompts that are tuned specifically to the functionality of the agents. 


\subsection{Matching Agent performance}

\begin{figure}[h]
    \centering
    \includegraphics[width=.45\textwidth]{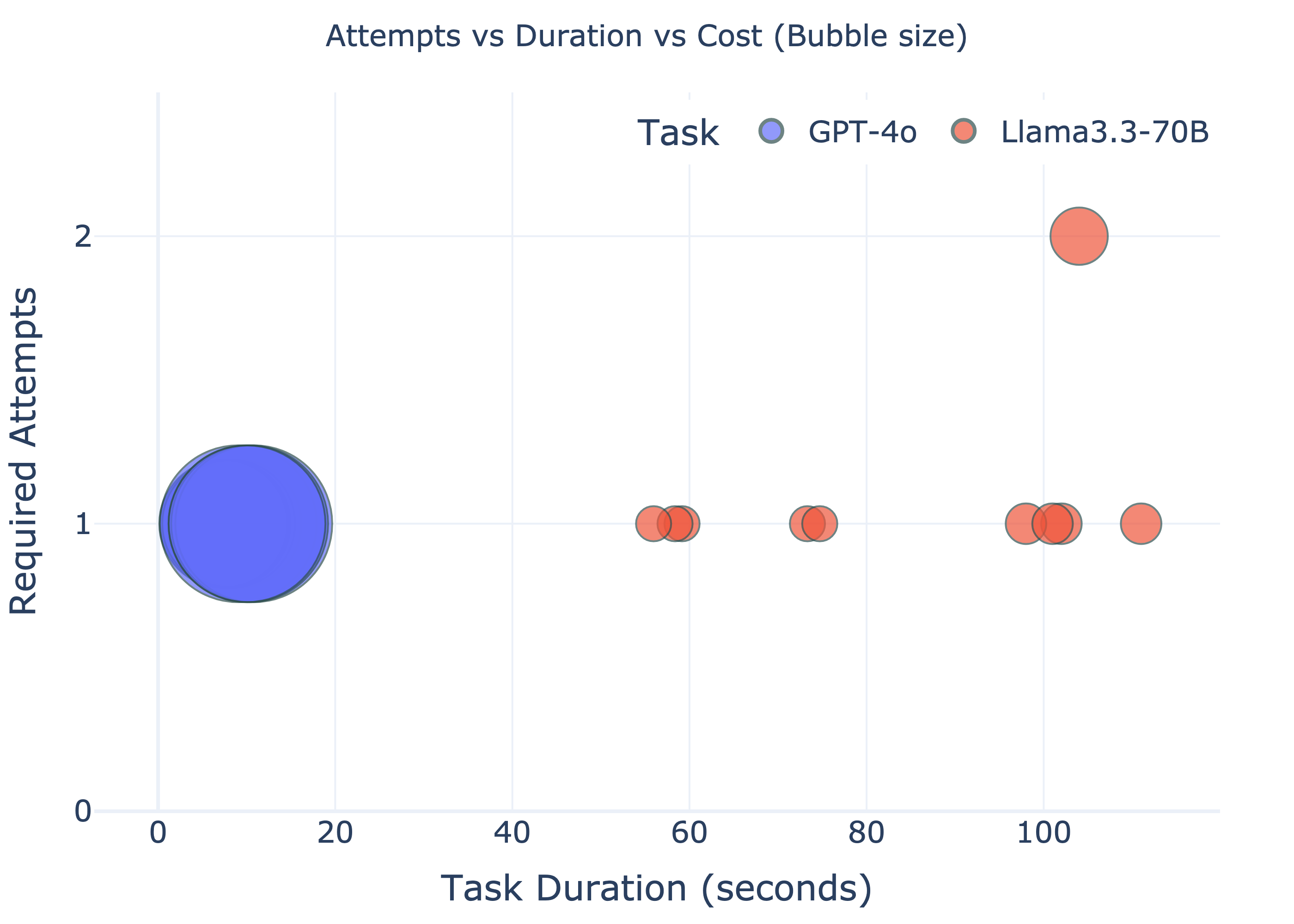} 
    \caption{Matching Agent performance}\label{fig:mat}
\end{figure}

To test the performance of the matching agent, we created 60 custom control function requirements based on the capability of AP/gNB. To account for the diverse nature of user input requirements, we generated ten distinct variations of these requirements using Claude 3.5 Sonnet\cite{claude}. We then used these variations as input to the matching agent and observed whether it correctly matches them to the capabilities of the AP/gNB and how many attempts were required to achieve a correct match. Figure~\ref{fig:mat} shows the number of attempts and the total time required to match 60 requirements. The diameter of the circle represents the cost. We observe that GPT-4o successfully matches all requirements on the first attempt. Llama3.3 achieves the same but with a few exceptions. Moreover, GPT-4o is 6 to 10 times quicker than Llama3.3, depending on the complexity of the matching task. However, GPT4-o is approximately 14 times more costly than Llama3.3.
\subsection{Codegen agent performance}
\begin{figure}[h]
    \centering
    \includegraphics[width=.45\textwidth]{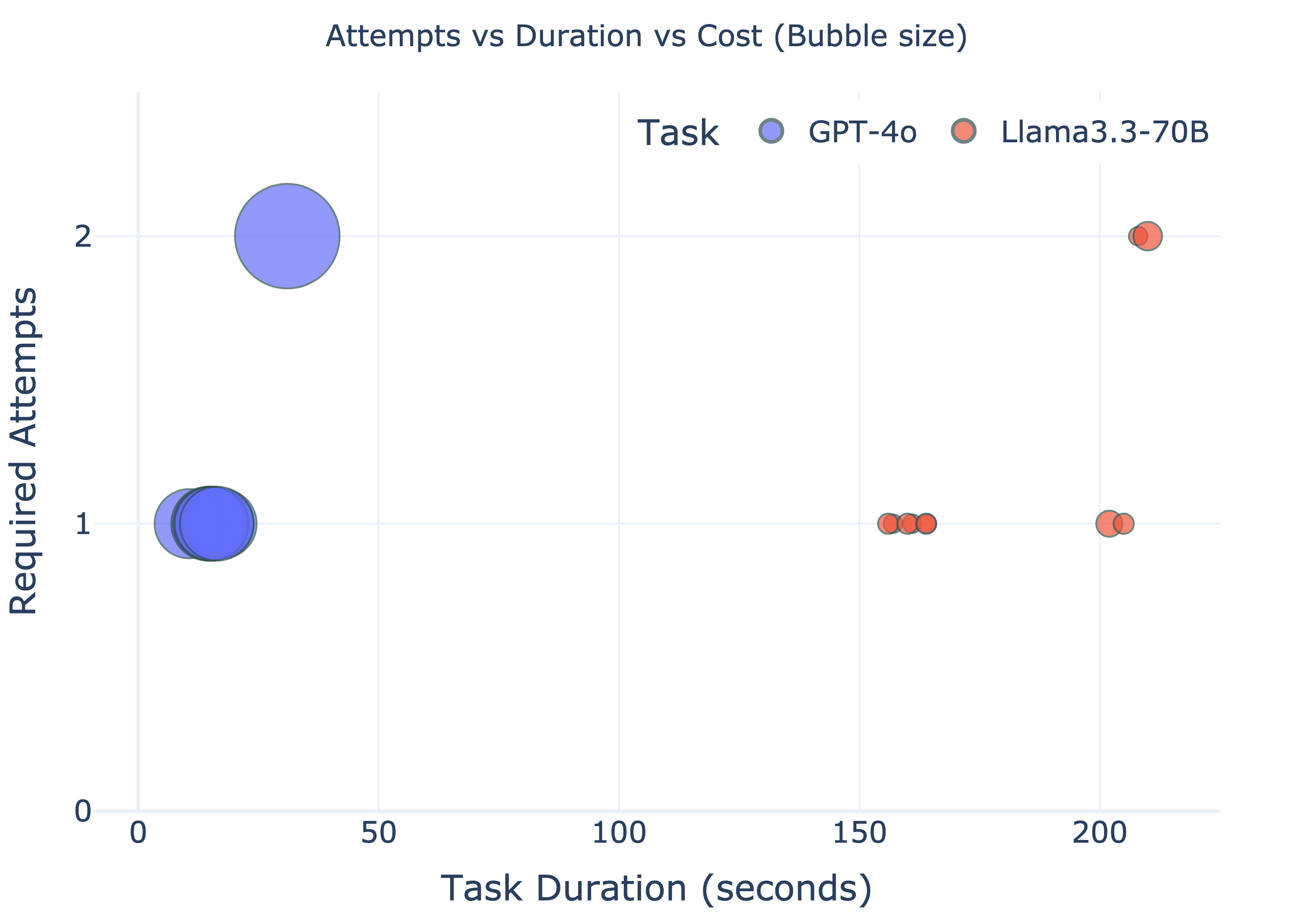} 
    \caption{Codegen agent performance}\label{fig:ca}
\end{figure}

To evaluate the performance of the codegen agent, we attempt to generate correct and executable code based on the CFR document provided by the matching agent. We generate code for 5 APs and 5 gNBs, each using distinct simulated vendor APIs. Figure~\ref{fig:ca} shows the number of attempts and the total time required to generate the code while the diameter of the circle represents the cost. We observe that the code generation was predominantly successful on the first attempt. 
Compared to the matching task, coding task is more complex. Here, the performance gap between Llama3.3 and GPT-4o increases significantly. In this case, GPT-4o based agent is 10-14 times faster compared to the Llama3.3 based agent but also costs 14 times more.

Besides the above experiments, we also tried to generate augmented control functions based on the existing NF capabilities. This approach is used when the control function requirement from the user does not have an exact match with the capabilities of the NF but a similar functionality is available. We tested two control functions:
\begin{itemize}
    \item Age of Information (AoI) aware rate control:  This control function is designed to set the specified data rate on the AP, provided that the given timestamp has not already elapsed. The rate setting is not applied if the given timestamp exceeds the AP's system time. In this case, the AP vendor supports rate control but lacks AoI aware rate control. The codegen agent is expected to add a wrapper on top of vendor API which can compare the provided timestamp against system time and generate the API server.
    \item Timestamped telemetry collection: In this case, the control function requires all the returned data to be timestamped. The gNB vendor supports data collection but does not include timestamps. The codegen agent is tasked with writing a wrapper around vendor API which can read system time and append it to returned data.
\end{itemize}

We discovered that the codegen agent was able to successfully augment the functions when clear and structured instructions were provided in the requirements. On average, this task took 3 times longer and was 3 times more expensive than generating non-augmented control functions. However, when we tried to create more complex augmented control functions involving generating multi-threaded code, the agent was unable to produce the correct code despite multiple attempts. This could only be resolved with manual intervention which involved modifying the original prompt.

\subsection{Overall results}

Overall, we observed that GPT4-o offered the best E2E performance. On average, generating an interface with 10 control functions costs \$0.04 and has an E2E latency of 9.4s. Furthermore, GPT-4 showed minimal variation in latency for a given agent with varying task complexity, whereas Llama3 shows a large variation in latency, even for similar kind of tasks.

\section{Limitations}

We highlight some of the limitations of our work below:
\begin{itemize}
    \item We conduct all of our experiments on synthetic data. This is due to the fact that most vendors do not expose their device API. Although we aim to model our synthetic data (control function requirements, vendor APIs) as realistically as possible, performance in real-world scenarios may vary significantly, necessitating further validation under diverse operational conditions. The performance of agents also depends on the availability of well-structured and detailed capability and API documentation\cite{Gunasekar2023}. 

    \item Considering real-world deployment of this framework, security of network functions might be affected by LLM specific threat vectors. These might be direct or indirect prompt injection or context windows overloading\cite{prompt_inject}. In our experiments, we observed that the size of the context window matters for accuracy. If the  number of control functions in a single query is too large the models start to produce inaccurate output\cite{netconfeval}. However, this can be addressed by splitting it into multiple smaller queries.
    \item In the case of matching agent, there is no ground truth against which the agent can validate the LLM model output. Hence, we had to validate the output by using another LLM for evaluation. This can be extended to use a set of specialized LLMs to evaluate the output which, however, also increases the cost. Moreover, this method may not be fully trustworthy due to the inherent uncertainty of LLMs. In real-world deployments, the generated interface might need to be tested in a sandbox before deploying in production.
    \item Due to the virtualization of the network edge, it is possible to deploy large AI models in the edge network functions. However, these models might be too big for devices such as WLAN APs. Although this could be mitigated by using cloud based models over an API, we believe that in the future the models would get small enough to be deployed on devices like APs.
\end{itemize}

\section{Related work}
NetconfEval\cite{netconfeval} introduces a model agnostic benchmark to evaluate LLMs for network configuration. It evaluates multiple use cases, developing routing algorithms, generating low level network configuration, converting high level requirements into formal specifications and function calls.

Netllmbench\cite{netllmbench} introduces a framework for benchmarking LLMs in network configuration tasks. It compares various LLMs including Llama3-70B in terms of their speed of operation and number of iterations required.

NAIL\cite{NAIL} discusses a method to translate high-level instructions from the user into executable code for P4 using transpiler. It also has ability to continuously monitor the network to ensure intent fulfillment.

The system COSYNTH~\cite{mondal_llm} uses verified prompt programming for generating router configurations. They demonstrate its application by translating Cisco router configurations into equivalent Juniper router configurations.

In \cite{WASM2024}, the authors discuss the integration and compatibility challenges in a multi-vendor O-RAN deployment. They propose a framework using WebAssembly plugins to enable interoperable and flexible multi-vendor deployments.

Zero-touch Service Management (ZSM)~\cite{LIYANAGE2022103362} frameworks can integrate AI/ML models to autonomously detect and resolve anomalies within 6G networks. 


\section{Conclusion and Future Work}

We present a novel framework to create network control interfaces on demand between any two set of NFs. We achieve this by leveraging LLMs to develop a multi-agent framework. Our framework brings flexibility and interoperability to the control interfaces. We believe that in future mobile networks, AI agents might be used natively in all parts of the network~\cite{huawei9}. Generating control interface on demand using AI is a step in this direction. 

In this work, we only focus on control interfaces and not on the applications that make the control decisions (e.g., xApps). We believe the decision-making applications can also be generated using AI in the future. We plan to investigate how generated control applications and interfaces could be used to create autonomous control loops for 6G networks.

\section*{Acknowledgment}
The authors acknowledge the financial support by the Federal Ministry of Education and Research of Germany (BMBF) in the program
of “Souverän. Digital. Vernetzt.” Joint project 6G-RIC (16KISK020K).
This project was also supported by German Federal Ministry for Digital and
Transport funded project 5G-COMPASS (19OI22017A).

\bibliographystyle{IEEEtran}
\bibliography{ref}

\end{document}